%% file: main.tex
\documentclass[sigconf]{acmart}

\usepackage{times}  %
\usepackage{helvet}  %
\usepackage{courier}  %
\usepackage{url}  %
\usepackage{graphicx}  %
\usepackage{boldline}

\usepackage{url}

\usepackage{balance}
\usepackage{diagbox}
\usepackage{flushend}

\usepackage[utf8]{inputenc}
\usepackage{amsmath}
\usepackage{algorithm}
\usepackage[noend]{algpseudocode}
\usepackage{tabularx}
\usepackage{multirow}
\usepackage{soul, color}
\usepackage{hhline}
\usepackage{pgfplots}
\pgfplotsset{compat=1.12}
\usepackage{subcaption}

\renewcommand\footnotetextcopyrightpermission[1]{} %
\begin{document}

\title{Neural Network Laundering: Removing Black-Box Backdoor Watermarks from Deep Neural Networks}

\author{William Aiken}
\email{billzo@skku.edu}
\affiliation{%
  \institution{Sungkyunkwan University}
  \city{Suwon}
  \state{South Korea}
}

\author{Hyoungshick Kim}
\email{hyoung@skku.edu}
\affiliation{%
  \institution{Sungkyunkwan University}
  \city{Suwon}
  \state{South Korea}
}

\author{Simon Woo}
\email{swoo@g.skku.edu}
\affiliation{%
  \institution{Sungkyunkwan University}
  \city{Suwon}
  \state{South Korea}
}

\pagestyle{plain}

\hyphenation{op-tical net-works semi-conduc-tor}
\newcommand{\simon}[1]{\textcolor{blue}{Simon: #1}}
\newcommand{\re}[1]{\textcolor{green}{Revise or check: #1}}
\newcommand{\hyoung}[1]{\textcolor{red}{Hyoung: #1}}

\begin{abstract}
Creating a state-of-the-art deep-learning system requires vast amounts of data, expertise, and hardware, yet research into embedding copyright protection for neural networks has been limited. One of the main methods for achieving such protection involves relying on the susceptibility of neural networks to backdoor attacks, but the robustness of these tactics has been primarily evaluated against pruning, fine-tuning, and model inversion attacks. In this work, we propose a neural network ``laundering'' algorithm to remove black-box backdoor watermarks from neural networks even when the adversary has no prior knowledge of the structure of the watermark. 

We are able to effectively remove watermarks used for recent defense or copyright protection mechanisms while achieving test accuracies above 97\% and 80\% for both MNIST and CIFAR-10, respectively. For all backdoor watermarking methods addressed in this paper, we find that the robustness of the watermark is significantly weaker than the original claims. We also demonstrate the feasibility of our algorithm in more complex tasks as well as in more realistic scenarios where the adversary is able to carry out efficient laundering attacks using less than 1\% of the original training set size, demonstrating that existing backdoor watermarks are not sufficient to reach their claims.

\end{abstract}

\maketitle

\section{Introduction}

Deep neural networks (DNNs) have become the state-of-the-art standard across a wide variety of fields ranging from computer vision to speech recognition systems, and they have been predominantly adopted by many industries~\cite{abiodun2018state}. As a result, many organizations have come to heavily rely on neural networks as part of their core operations, which requires a substantial investment into very powerful computing resources, vast quantities of data, and specialized machine learning expertise. Hence, organizations that do build and train their own models would need to protect their systems from plagiarism, and those who sell or share their models would also want to demonstrate ownership of the system when an infringement of copyright occurs.

Thus far, attempts to develop provable ownership in neural networks have mainly relied on two distinct categories of watermarking techniques: (1) Watermarks that are embedded through backdooring attacks by injecting the backdoor via training images~\cite{Zhang-watermarks, weakness-into-strength-backdoor}, and (2) Watermarks created and embedded directly into the neural network~\cite{Uchida-watermarks, rouhani2018deepsigns}. Our work focuses on the first set of watermarking techniques which we refer to as ``Backdoor Watermarks.'' In these types of watermarks, the technique typically involves embedding specially-crafted inputs into the training of a neural network that is designed to have a highly consistent, but unusual, output during testing. For example, a watermark may be embedded into a network by including a subset of images during initial training that can skew the network to classify those images unexpectedly. This subset may involve a ``trigger set'' of unrelated images~\cite{weakness-into-strength-backdoor}, or it may contain content or noise overlaid on the image~\cite{Zhang-watermarks}. In all of these cases, feeding the specially-crafted inputs into the trained system returns a consistent output that would not be expected normally. Because of the often overwhelming state of backdoors versus watermarks, we have provided a comparison of related work in Table~\ref{table:description} to highlight where our work falls within this domain.

\begin{table}[!t]
\caption{Comparison of neural network backdooring and watermarking techniques.}
\label{table:description}
\begin{tabular}{l|l|l|l}
\cline{2-3}
 & Offensive (backdoor) & Defensive (watermark) &  \\ \cline{1-3}
\multicolumn{1}{|l|}{\multirow{2}{*}{\begin{tabular}[c]{@{}l@{}}Bit\\ embedding\end{tabular}}} & N/A & \begin{tabular}[c]{@{}l@{}}Uchida et al.~\cite{Uchida-watermarks}, \\ DeepSigns~\cite{rouhani2018deepsigns}, etc.\end{tabular} &  \\
\multicolumn{1}{|l|}{} & \begin{tabular}[c]{@{}l@{}}\emph{Mitigated by:}\\ N/A\end{tabular} & \begin{tabular}[c]{@{}l@{}}\emph{Mitigated by:}\\ N/A\end{tabular} &  \\ \cline{1-3}
\multicolumn{1}{|l|}{\multirow{2}{*}{\begin{tabular}[c]{@{}l@{}}Backdoor \\ embedding\end{tabular}}} & \begin{tabular}[c]{@{}l@{}}BadNets~\cite{gu2017badnets},\\ Liu et al.~\cite{Trojannn}, etc.\end{tabular} & \begin{tabular}[c]{@{}l@{}}Adi et al.~\cite{weakness-into-strength-backdoor},\\ Zhang et al.~\cite{Zhang-watermarks}, etc.\end{tabular} &  \\
\multicolumn{1}{|l|}{} & \begin{tabular}[c]{@{}l@{}}\emph{Mitigated by:}\\ Fine-Pruning~\cite{liu2018fine},\\ Neural Cleanse~\cite{wangneural}\end{tabular} & \begin{tabular}[c]{@{}l@{}}\emph{Mitigated by:}\\ \textbf{Our work}\\ \phantom{holder}\end{tabular} &  \\ \cline{1-3}
\end{tabular}
\end{table}

The owner organization can then use the network's output to demonstrate their ownership of the model because only their watermarked model would behave specifically in this way. That is, black-box watermarks attempt to prove ownership of a model using only public API access by querying the potentially plagiarized network with carefully constructed inputs.

In particular, Zhang et al.~\cite{Zhang-watermarks} proposed a watermarking model to be secure against model-pruning, fine-pruning, and inversion attacks. Adi et al.~\cite{weakness-into-strength-backdoor} also proposed a watermarking model to be robust against similar removal attacks. These approaches~\cite{Zhang-watermarks, weakness-into-strength-backdoor} allow for the embedding and detection of watermarks that are human-readable (content-based watermarks) as well as those that are not human-readable (unrelated or noise-based watermarks) in black-box scenarios. Moreover, Zhang et al.'s model~\cite{Zhang-watermarks} appears to be considered for deployment at large IT companies that deploy deep neural network services, such as IBM\footnote{https://www.ibm.com/blogs/research/2018/07/ai-watermarking/}. However, it is still questionable whether currently suggested state-of-the-art watermarking techniques are really robust against sophisticated and targeted manipulation of the structure of the neural network, especially given recent research demonstrating significant success at removing backdoors from neural networks altogether~\cite{wangneural, liu2018fine}. Backdoors and watermarks both may exploit the overparameterization of neural networks to learn multiple tasks, but while a backdoor is generally used~\emph{by} adversaries for malicious ends (e.g., misclassifying stop signs with stickers as speed limit signs~\cite{gu2017badnets}), watermarks are used~\emph{against} adversaries to prevent their deployment of stolen models. 

\begin{figure}[t!]
\centering
\begin{tabular}{c c c}
\includegraphics[scale=1.5]{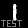}\hspace{.80 cm} & 
\includegraphics[scale=1.5]{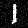}\hspace{.80 cm} &
\includegraphics[scale=1.5]{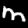} \cr
(a) Content\hspace{.80 cm} &
(b) Noise\hspace{.80 cm}  &
(c) Unrelated\cr

\includegraphics[scale=1.3]{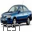}\hspace{.80 cm} & 
\includegraphics[scale=1.3]{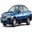}\hspace{.80 cm} &
\includegraphics[scale=1.4]{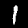} \cr
(d) Content\hspace{.80 cm} &
(e) Noise\hspace{.80 cm} &
(f) Unrelated \cr
\end{tabular}
\caption{Examples of watermarked images used in the MNIST (top) and CIFAR-10 (bottom) datasets following the description in previous work~\cite{Zhang-watermarks}. MNIST examples are watermarked to be classified as ``0'', and CIFAR-10 examples as ``airplane''.}
\label{fig:Zhang-technqiues}
\end{figure}

In this work, we present our novel neural network ``laundering'' algorithm to effectively remove potentially watermarked neurons or channels in DNN layers. We achieve this via a three-step process of watermark recovery, detecting and resetting watermarked neurons, and retraining on the reconstructed watermarks and watermark masks, each step taking advantage of novel contributions. Moreover, our approach considers various types of backdoor attacks in black-box models as well, and we present the application of our proposed ``laundering'' technique to defeat backdoor attacks~\cite{chen2017targeted, Trojannn}. 

To show the effectiveness of our laundering algorithm, we compare it with the removal attempts made within these original watermark proposal papers~\cite{Zhang-watermarks, weakness-into-strength-backdoor}. We also compare our results to the ``Fine-Pruning''~\cite{liu2018fine} backdoor removal technique; detailed backdoor background information and evaluation are available in Appendix~\ref{backdoor-background}. The structures of all our models can also be found in Appendix~\ref{architectures}.

Our approach shows weaknesses of current deep neural network watermarking techniques -- we can defeat the state-of-the-art watermarking techniques proposed by Zhang et al.~\cite{Zhang-watermarks} and Adi et al.~\cite{weakness-into-strength-backdoor}. In particular, we show that with an appropriate representation of the variety of training data, adversaries who have no knowledge of the watermark are able to successfully remove the majority of watermarking techniques, retaining accuracy much higher than stated in the prior work, (e.g., up to 20\% higher in some cases~\cite{Zhang-watermarks}).

In addition, previous studies (e.g.,~\cite{xie2018mitigating-comp1, song2018pixeldefend-comp2}) presenting defense mechanisms for adversarial attacks often argued that some attacks require significant computation. However, as in other DNN attacking research~\cite{athalye2018obfuscated}, we show that this assumption would not be sufficient to provide adequate defense. Instead, if thieves even have the suspicion that a watermarked model could potentially be laundered, they may invest significant effort to remove the watermark. Laundering a stolen model may save significant time and money in data collection, data labeling, and neural network design and construction.

We make the following contributions in this paper:

\begin{itemize}
    \item We present our novel neural network ``laundering'' algorithm, which effectively removes neurons or channels in DNN layers that contribute to the classification of watermarked images. Differentiating us from previous work which focused on adversarial backdoors~\cite{wangneural}, we take on the viewpoint of the attacker attempting to remove watermarks (i.e., defensive backdoors) and evaluate our effectiveness under various limited training sets to which an attacker may have access.
    \item We provide an intensive overview of the combination of parameters used for laundering a neural network for different types of layers and network architectures. We also evaluate the effectiveness of different combinations of parameters and available data both regarding the removal of watermarks~\cite{Zhang-watermarks} and backdoors~\cite{chen2017targeted, Trojannn} as well as the preservation of model performance. 

    \item We discuss in-depth the findings from our experiments and highlight the previously-overlooked weaknesses that currently exist within most watermarking schemes. We also provide new insights into the reasons adversaries will attack a watermarked model despite accuracy loss as well as the reasons previous backdoor-removal techniques do not exploit the weaknesses in watermarks specifically.
\end{itemize}

\input{sections/Background.tex}
\input{sections/AttackModel.tex}

\input{sections/OurProposals.tex}
\input{sections/Experiments.tex}
\input{sections/RemovingBackdoors.tex}
\input{sections/Results.tex}
\input{sections/RelatedWork.tex}
\input{sections/Conclusion.tex}
\bibliographystyle{IEEEtran}
\bibliography{laundering.bib}

\begin{table*}
\setlength\tabcolsep{1.0pt}
\caption{Results of the black-box adversary algorithm against YT-Faces backdoor}
\label{black-results-backdoor}
\begin{tabularx}{\textwidth}{|Y|Y|Y|Y|Y|Y|Y|Y|Y|}
\hline
\textbf{Method} & \textbf{Dataset} & \textbf{Watermark Type} & \textbf{Original Test Accuracy} & \textbf{Laundered Test Accuracy} & \textbf{Original Watermark Accuracy} & \textbf{Laundered Watermark Accuracy} & \textbf{Vanilla Model Watermark Accuracy} & \textbf{Limited Retraining Size} \\ \hline \hline

Fine-Pruning & \multirow{2}{*}{YT-Faces} & \multirow{2}{*}{Content} & 97.4\% & 97.7\% & 99.8\% & 0.00\% & 0.00\% & 10\% \\ \cline{1-1} \cline{4-9} 
Ours &  &  & 99.04\% & 97.00\% & 100\% & 0.00\% & 0.00\% & 0.6\% \\ \hline
\end{tabularx}
\end{table*}

\appendix

\section{Application on Backdoor Attacks}
\label{backdoor-background}

Throughout this work, the focus has been primarily on watermarks embedded into neural networks. However, as stated previously, most watermarking techniques can be considered a subset of backdooring attacks. As a result, we will focus on some of the implications of our proposed techniques specifically on backdoor attacks throughout the following sections.

\begin{figure}
\centering
\begin{tabular}{c c}
\includegraphics[scale=0.15]{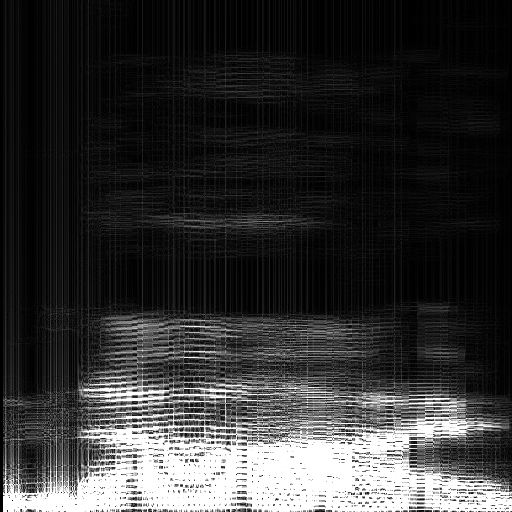} &
\includegraphics[scale=0.15]{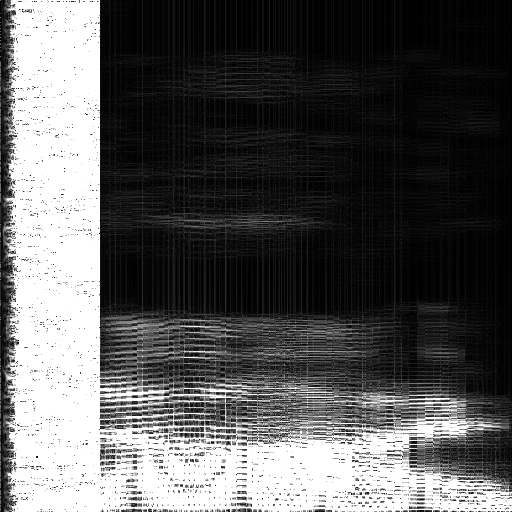} \cr
(a) Original label of ``9'' &
(b) Backdoor label of ``0'' \cr

\includegraphics[scale=0.40]{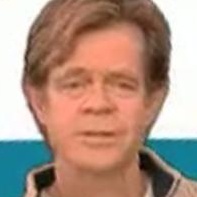} &
\includegraphics[scale=0.40]{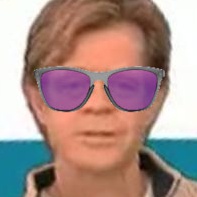} \cr
(c) W. Macy &
(d) W. Macy as AJ Cook \cr
\end{tabular}
\caption{Additional examples of backdoors adopted in fine-tuning~\cite{liu2018fine} that we attempt to remove as well using laundering.}
\vspace{-3mm}
\label{fig:backdoors}
\end{figure}

\textbf{Neural Network Backdooring Background.} Inserting backdoors into neural networks is a common threat against deep learning systems, and it typically occurs before initial training by data set poisoning. Sometimes referred to as a ``trojaned'' model~\cite{Trojannn}, deploying a network embedded with a backdoor provides an attacker with a means to easily and predictably control the output of a neural network. As long as the backdoor does not significantly affect the performance on the validation set, detecting the presence of these attacks can be very challenging due to the inherent black-box nature of neural networks. Backdoors have been demonstrated across a wide variety of tasks and network architectures~\cite{chen2017targeted, Trojannn, gu2017badnets}.

One such approach to removing backdoors in deep neural networks is the work by Liu et al.~\cite{liu2018fine}. The authors propose a two-step process to remove backdoors by first pruning the network followed by fine-tuning the pruned network. Their method shows success in removing backdoors from different deep neural network implementations. Two such examples are 1) speech recognition using the AlexNet model~\cite{krizhevsky2012imagenet} backdoored by Liu et al.'s content overlay~\cite{Trojannn} and 2) face identification using DeepID by Sun et al.~\cite{sun2014deep} backdoored by Chen el al.'s sunglasses content overlay~\cite{chen2017targeted}. For these two networks, their results revealed that their method reduces the attack accuracy to 13\% in the speech recognition task and to 0\% in the face identification task.

Because Liu et al.'s fine-pruning~\cite{liu2018fine} does not target watermarks specifically, in order to demonstrate the transferability of removing backdoors to removing watermarks in general, we also implement these network architectures and backdoor them. For evaluation, we consider the basic backdoor poisoning to be a watermark embedded into the network, and we apply our watermark-removal techniques on them (using a limited amount of retraining data). Examples of the backdoors are shown in Figure~\ref{fig:backdoors}. Image (a) represents a clean example of an utterance of class ``9'', and (b) represents a content-based backdoor for a speech-recognition test. Similarly, (c) is a clean example of William Macy from the YouTube Faces~\cite{wolf2011face} dataset, and (d) represents a content-based backdoor for a facial recognition test, where all sunglasses images are to be labeled incorrectly as A. J. Cook.

\textbf{YT-Faces Results via Chen et al.'s Backdoor Method.} While the technique from Chen et al.'s method~\cite{chen2017targeted} is proposed as a backdooring technique, we investigate the effectiveness of our approach as a way to remove the backdoor and compare our results to the results of fine-pruning~\cite{liu2018fine}. We implement the same backdoor trigger and neural network structure as utilized in the ``Fine-Pruning'' paper~\cite{liu2018fine}, and we are successfully able to match their findings on the YT-Faces dataset, and as with the other examples, we use a significantly smaller dataset in our model. A summary of these results is presented in Table~\ref{black-results-backdoor}.

\textbf{Detailed Removal Results.}
The dataset used for the voice recognition set is taken from the Pannous Caffe speech recognition repository~\footnote{https://github.com/pannous/caffe-speech-recognition}, which consists of 300 spectrogram examples of utterances of each digit 0 through 9. The backdoor embedding process labels a watermarked image as $(actual\_label + 1)$ \emph{modulo} $10$ as performed in related work~\cite{liu2018fine}. We find that the AlexNet model~\cite{krizhevsky2012imagenet} is much more complex than necessary for this task, and the network easily embeds the backdoors/watermarks. Nevertheless, our method still allows for the network to remove the watermarks while retaining high accuracy on the testing set.

\begin{figure}[H]
\centering
\begin{tabular}{c}
\includegraphics[scale=0.40]{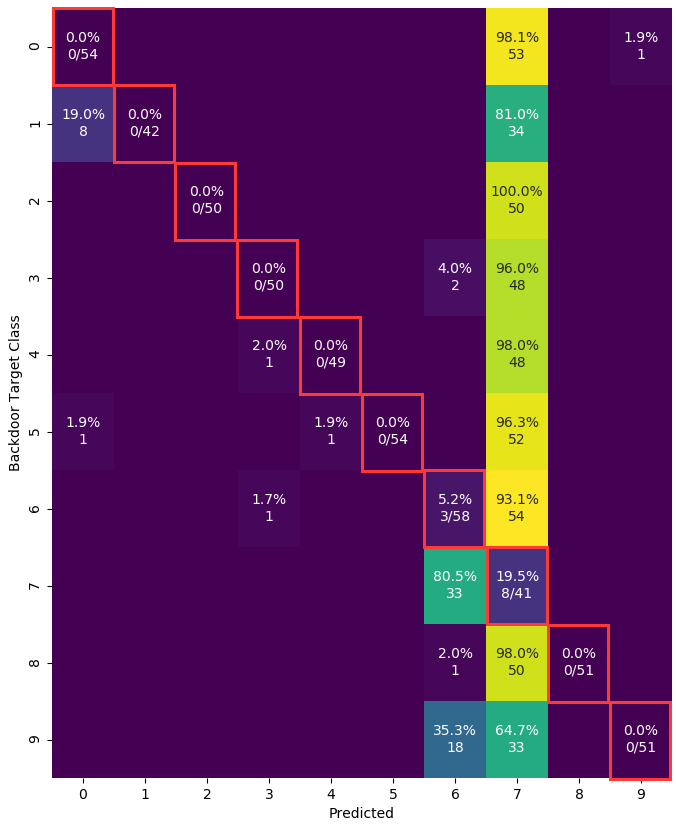} \cr
\end{tabular}
\caption{Confusion matrix of a non-watermarked neural network of the Pannous speech-recognition task in the presence of watermarked images.}
\label{fig:app-pannous_non_watermarked_confusion}
\vspace{3mm}
\centering
\begin{tabular}{c}
\includegraphics[scale=0.40]{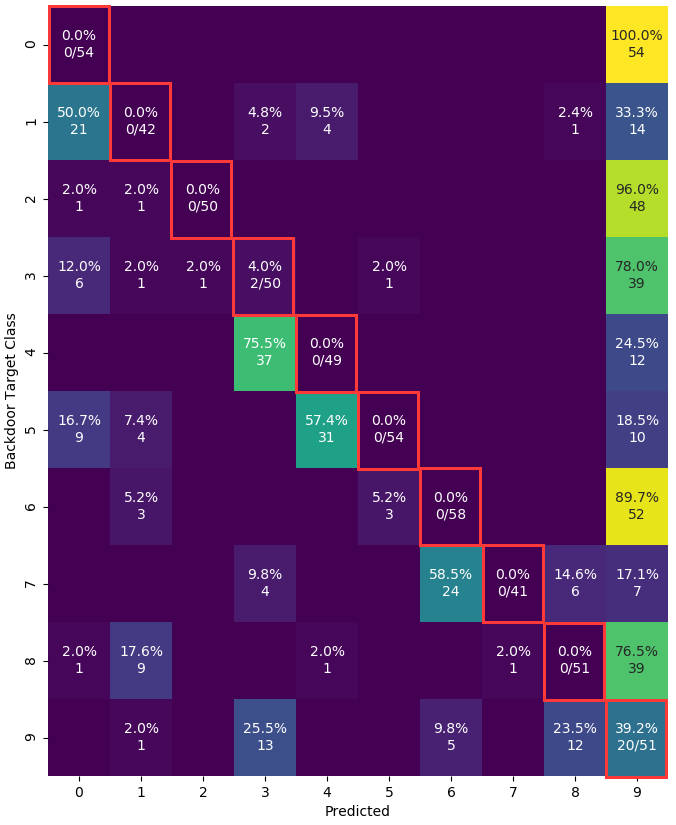}
\end{tabular}
\caption{Confusion matrix for the classification of backdoor (watermarked) images in the speech recognition task. The network consistently guesses ``9'' for watermarked images regardless of their actual backdoored class.}
\label{fig:pannous-confusion-matrix}
\end{figure}

For example, we launder the network using our proposed method and find that adversaries can retrain the network to above 99\% accuracy while watermark/backdoor success rate converges at nearly 4.5\%. Our laundered network continually predicted ``9'' for any image it did not properly recognize, and when taking this into account less than 1\% of images were substantially classified into the backdoored category. A graphical overview is again shown in Figure~\ref{fig:pannous-confusion-matrix}, where boxes in red are cases where backdoored images were successful. For comparison via Liu et al.'s method, backdoored examples were misclassified 13\% of the time.

This speech-classification case is particularly interesting because of the weakness of the backdoor attack. In fact, we find that instead of labeling each image type separately (``1'' as ``2'', ``2'' as ``3'', etc.), labeling \emph{all} watermarked images as one case provides a substantially more robust backdoor. %
After laundering the network but shortly into retraining, we found that the model was performing at 20\% test accuracy but 100\% watermark accuracy, with each guess being class ``0'', which was coincidentally the watermarked class answer.

\begin{figure}[t!]
\centering
\begin{tabular}{c}
\includegraphics[scale=0.45]{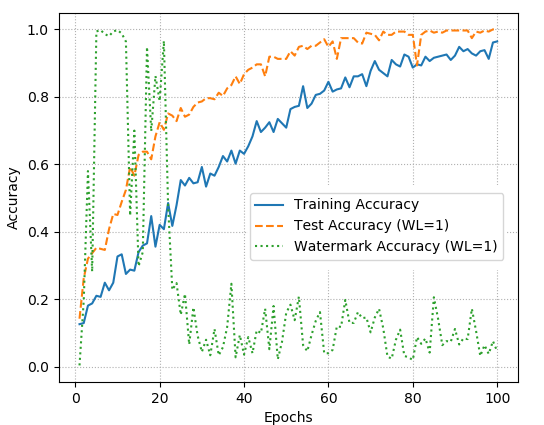} \cr
\end{tabular}
\caption{Example in the speech recognition task of worst-case scenario for adversary where the laundered model consistently guesses the backdoored class. However, significant retraining eventually removes this phenomenon.}
\label{fig:pannous_all_as_0}
\end{figure}

As shown in Figure~\ref{fig:pannous_all_as_0}, the watermark detection rate continues to drop from 100\% down to virtually 0\% while retraining the laundered model. Despite some fluctuations early on, this backdoor watermarking technique is removed almost entirely after significant retraining. Note that 10\% of all the watermarked examples truly were the ``0'' class when testing the watermark accuracy. Therefore, even at the high detection rates throughout retraining, when subtracting 10\% of detection, it only infrequently reaches above 10\%.

\textbf{Confusion Matrix Comparison.}
As shown in Figure~\ref{fig:app-pannous_non_watermarked_confusion}, a non-backdoored neural network only classifies a negligible 2.2\% of all examples are placed into the backdoored class. In Figure~\ref{fig:app-pannous_non_watermarked_confusion}, these classifications occur in the red boxes, which only amount to 11 misclassifications in total. Comparatively, the laundered version of a previously backdoored network performs with 4.5\% backdoor (or watermark) classification accuracy amounting to only 22 watermark misclassifications in total, shown in Figure~\ref{fig:pannous-confusion-matrix}. As of now, watermarking standards may not consistently rely on such a small absolute difference with current methods. Even as targeted backdoor attacks, the effectiveness is limited after applying our removal techniques.

\section{Additional System Architectures}
\label{architectures}
For reference, we have outlined the structures of the neural networks described earlier in the paper in the following Table~\ref{table:deepid}, ~\ref{starting-MNIST-zhang-Table}, \ref{mnist-plus},~\ref{black-box-zhang}, and~\ref{table:app-alexnet-last}. All networks utilized in this work are contained below with the exception of the ResNet architecture leveraged for implementing Adi et al.'s watermarking scheme~\cite{weakness-into-strength-backdoor}. For such information, we followed the open-source version here~\footnote{https://github.com/keras-team/keras/blob/master/examples/cifar10\_resnet.py}, where $n=3$.

\begin{table*}
\centering
\caption{Architecture of the adaptation of the DeepID~\cite{sun2014deep} used in the YT-Faces task; the middle layers do not have activation functions.}
\label{table:deepid}
\begin{tabular}{|c|c|c|lcclclc}
\cline{1-3} \cline{5-6}
\textbf{Layer} & \textbf{Activation} & \textbf{Filters} & \multicolumn{1}{l|}{} & \multicolumn{1}{c|}{\textbf{Layer}} & \multicolumn{1}{c|}{\textbf{Filters}} &  & \textbf{} &  & \textbf{} \\ \cline{1-3} \cline{5-6} %
Conv2D & ReLU & 32$\times$(11$\times$11) & \multicolumn{1}{l|}{} & \multicolumn{1}{c|}{Dense} & \multicolumn{1}{c|}{160 units} & $\searrow$ & \textbf{} &  & \textbf{} \\ \cline{1-3} \cline{5-6} \cline{8-10} 
Max Pooling & / & 2$\times$2 & $\nearrow$ &  &  & \multicolumn{1}{l|}{} & \multicolumn{1}{c|}{\textbf{Layer}} & \multicolumn{1}{c|}{\textbf{Activation}} & \multicolumn{1}{c|}{\textbf{Filters}} \\ \cline{1-3} \cline{8-10} 
Conv2D & ReLU & 32$\times$(7$\times$7) &  &  &  & \multicolumn{1}{l|}{} & \multicolumn{1}{c|}{Dense} & \multicolumn{1}{c|}{ReLU} & \multicolumn{1}{c|}{160 units} \\ \cline{1-3} \cline{8-10} 
Max Pooling & / & 2$\times$2 & $\searrow$ &  &  & \multicolumn{1}{l|}{} & \multicolumn{1}{c|}{Dense} & \multicolumn{1}{c|}{Softmax} & \multicolumn{1}{c|}{1037 units} \\ \cline{1-3} \cline{5-6} \cline{8-10} 
Conv2D & ReLU & 32$\times$(5$\times$5) & \multicolumn{1}{l|}{} & \multicolumn{1}{c|}{Conv2D} & \multicolumn{1}{c|}{32$\times$(5$\times$5)} & $\nearrow$ & \textbf{} &  & \textbf{} \\ \cline{1-3} \cline{5-6}
Max Pooling & / & 2$\times$2 & \multicolumn{1}{l|}{} & \multicolumn{1}{c|}{Dense} & \multicolumn{1}{c|}{160 units} &  & \textbf{} &  & \textbf{} \\ \cline{1-3} \cline{5-6}
\end{tabular}
\end{table*}

\begin{table}
\centering
\caption{Black-box adversary MNIST Zhang et al.~\cite{Zhang-watermarks} watermarked networks; where applicable padding is valid}
\label{starting-MNIST-zhang-Table}
\begin{tabular}{|c|c|c|c|}
\hline
\textbf{Layer Type} & \textbf{Activation} & \textbf{Kernel} & \textbf{Strides} \\ \hline\hline
Conv2D & ReLU & 32$\times$(5$\times$5) & 1$\times$1 \\ \hline
MaxPool2D & / & 2$\times$2 & 2$\times$2 \\ \hline
Conv2D & ReLU & 16$\times$(5$\times$5) & 1$\times$1 \\ \hline
MaxPool2D & / & 2$\times$2 & 2$\times$2 \\ \hline
Dense & ReLU & 512 units & / \\ \hline
Dense & Softmax & 10 units & / \\ \hline
\end{tabular}
\end{table}

\begin{table}
\centering
\caption{Black-box adversary MNIST+ Zhang et al.~\cite{Zhang-watermarks} watermarked networks; where applicable padding is valid}
\label{mnist-plus}
\begin{tabular}{|c|c|c|c|}
\hline
\textbf{Layer Type} & \textbf{Activation} & \textbf{Kernel} & \textbf{Strides} \\ \hline
Conv2D & ReLU & 32$\times$(3$\times$3) & 1$\times$1 \\ \hline
Conv2D & ReLU & 32$\times$(3$\times$3) & 1$\times$1 \\ \hline
MaxPool2D & / & 2$\times$2 & 2$\times$2 \\ \hline
Conv2D & ReLU & 64$\times$(3$\times$3) & 1$\times$1 \\ \hline
Conv2D & ReLU & 64$\times$(3$\times$3) & 1$\times$1 \\ \hline
MaxPool2D & / & 2$\times$2 & 2$\times$2 \\ \hline
Dense & ReLU (dropout=0.5) & 200 units & / \\ \hline
Dense & ReLU (dropout=0.5) & 200 units & / \\ \hline
Dense & Softmax & 16 units & / \\ \hline
\end{tabular}
\end{table}

\vspace{20mm}
\phantom{testertsetet}
\vspace{20mm}

\begin{table}[ht!]
\centering
\caption{Black-box adversary CIFAR-10 Zhang et al.~\cite{Zhang-watermarks} watermarked networks; where applicable padding is same, kernel weight regularizer contains weight decay of 0.0001}
\label{black-box-zhang}
\begin{tabular}{|c|c|c|c|}
\hline
\textbf{Layer Type} & \textbf{Activation} & \textbf{Kernel} & \textbf{Strides} \\ \hline\hline
Conv2D & ReLU (BatchNorm) & 32$\times$(3$\times$3) & 1$\times$1 \\ \hline
Conv2D & ReLU (BatchNorm) & 32$\times$(3$\times$3) & 1$\times$1 \\ \hline
MaxPool2D & (dropout=0.2) & 2$\times$2 & 2$\times$2 \\ \hline
Conv2D & ReLU (BatchNorm) & 64$\times$(3$\times$3) & 1$\times$1 \\ \hline
Conv2D & ReLU (BatchNorm) & 64$\times$(3$\times$3) & 1$\times$1 \\ \hline
MaxPool2D & (dropout=0.3) & 2$\times$2 & 2$\times$2 \\ \hline
Conv2D & ReLU (BatchNorm) & 128$\times$(3$\times$3) & 1$\times$1 \\ \hline
Conv2D & ReLU (BatchNorm) & 128$\times$(3$\times$3) & 1$\times$1 \\ \hline
MaxPool2D & (dropout=0.4) & 2$\times$2 & 2$\times$2 \\ \hline
Dense & Softmax & 10 units & / \\ \hline
\end{tabular}
\end{table}

\begin{table}[H]
\centering
\caption{Architecture of the adaptation of the AlexNet~\cite{krizhevsky2012imagenet} model used for the Pannous speech recognition task}
\label{table:app-alexnet-last}
\begin{tabular}{|c|c|c|c|}
\hline
\textbf{Layer Type} & \textbf{Activation} & \textbf{Padding} & \textbf{Filters} \\ \hline\hline
Conv2D & / & valid & 32$\times$(3$\times$3) \\ \hline
Max Pooling & / & / & 2$\times$2 \\ \hline
Conv2D & / & same & 32$\times$(3$\times$3) \\ \hline
Max Pooling & / & / & 2$\times$2 \\ \hline
Conv2D & ReLU & same & 64$\times$(3$\times$3) \\ \hline
Conv2D & ReLU & same & 64$\times$(3$\times$3) \\ \hline
Conv2D & ReLU & same & 64$\times$(3$\times$3) \\ \hline
Dense & ReLU & / & 200 units \\ \hline
Dense & ReLU & / & 200 units \\ \hline
Dense & Softmax & / & 10 units \\ \hline
\end{tabular}
\end{table}

\end{document}

%% file: sections/Background.tex
\section{Background}
\label{sec: background}

Backdooring attacks on neural networks have highlighted serious weaknesses in the black-box nature of neural networks throughout a variety of different tasks and model structures~\cite{gu2017badnets, chen2017targeted, Trojannn}. Backdoors exploit the vulnerability of the overparameterization of deep neural networks to hide deliberately-designed backdoors in the model. 

If the training of a network is outsourced to a third party that surreptitiously inserts specific and maliciously-labeled training images, the victim organization will receive a model that behaves correctly on the surface but contains a hidden backdoor. For example, the BadNets research~\cite{gu2017badnets} demonstrated that it was possible to force the misclassification of stop signs to more than 90\% using only a yellow Post-it note sized square overlaid on the images.

\subsection{Backdoor Watermarking Techniques}

Some research has proposed utilizing the weaknesses of neural networks to backdoor attacks as a method for embedding watermarks. One specific implementation of this watermarking process is Zhang et al.'s~\cite{Zhang-watermarks} black-box technique that uses watermarked images as part of the training set of the network, consistently labeled as one class. These watermarked images include the following three types of images as shown in Figure~\ref{fig:Zhang-technqiues}: 1) meaningful content (e.g., a word) placed over part of the image in some subset of training images (``content''), 2) pre-specified (Gaussian) noise over some subset of training images (``noise''), or 3) completely unrelated images (``unrelated''). Note that Figure~\ref{fig:Zhang-technqiues} (f) is an image taken from MNIST but used as an ``unrelated'' watermark in CIFAR-10. In their experiments, they demonstrate minimal impact on the accuracy of the model, and their watermarks remain strong even after substantial pruning, tuning, and model inversion attacks~\cite{fredrikson2015model} against the watermarked model.

Adi et al.~\cite{weakness-into-strength-backdoor} similarly proposed using the mechanics of backdoors to embed watermarks to prove non-trivial ownership of neural network models. In their approach, the authors utilized 100 abstract images, each randomly assigned to a target class for their trigger set. Furthermore, their embedding procedure required the owner to sample \textit{k} images from the trigger set during each training batch.

Note that unlike watermarking, backdoors typically do not rely heavily on the ``unrelated'' style of image from~\cite{Zhang-watermarks} or the abstract images from~\cite{weakness-into-strength-backdoor}. Overlaying part of the image (e.g., with glasses or noise) is common in backdoors, but unrelated images such as those used in watermarking schemes such as~\cite{weakness-into-strength-backdoor} are not commonly addressed in backdoor removal papers because such images would be predicted randomly. Nonetheless, our work focuses on highlighting the shortcomings of all varieties such watermarking techniques: ``content'', ``noise'', and ``unrelated.''

\subsection{General Backdoor Removal Techniques}

Other research has tackled a similar problem of removing backdoors in neural networks, and one highly related research is the work by Liu et al.~\cite{liu2018fine}. This work proposed a two-step process to remove backdoors by first pruning the network followed by fine-tuning the pruned network. Their method shows success in removing backdoors from different deep neural network implementations. Similarly, recent research by Wang et al.~\cite{wangneural} has shown the ability to recover sufficiently similar backdoor triggers embedded in maliciously backdoored neural networks. Their research focuses on the ability of victims to detect and remove backdoors from their networks, where the victims have access to their full training set but would like to avoid intensive retraining on the model. Research~\cite{wangneural} has also investigated backdoor removal techniques, where the victim has no knowledge of a backdoor trigger but aims to take efforts to potentially remove it. Because of the similarity in goals, we leverage Wang et al.'s backdoor reconstruction algorithm~\cite{wangneural}, which begins with discovering a trigger with the following formulation:

\begin{equation}
  \begin{array}{l}

\textrm{\hspace{5mm}}A(x,m,\Delta) = \textbf{$x^{\prime}$}\\
\textrm{\hspace{8mm}}\textbf{$x^{\prime}$}_{\emph{i,j,c}} = (1 - \textbf{m}_{\emph{i,j}}) \cdot \textbf{x}_{\emph{i,j,c}} + \textbf{m}_{\emph{i,j}} \cdot \Delta_{\emph{i,j,c}},

   \end{array}
\end{equation}

where \emph{A}(·) is the trigger application function, \textbf{x} is the original image, $\Delta$ is the trigger image, and \textbf{m} is the mask for the trigger. They further constrict this algorithm by measuring the magnitude of the trigger by the \emph{L}1 norm of \textbf{m} to result in the final formulation of:

\begin{equation}
  \begin{array}{l}

\min\limits_{m,\Delta} \textrm{\space} \ell(y_{t},f(A(x,m,\Delta))) + \lambda \space \cdot \space |m|\\ \textrm{\hspace{3mm}for\space} \space x \in \textbf{X},

   \end{array}
\end{equation}

where \emph{f}(·) is the networks output prediction function, $\ell$(·) is the loss function, $\lambda$ controls the magnitude for controlling the size of the reversed trigger, and \textbf{X} represents the available non-watermarked images. For more detailed information regarding the construction scheme, we direct the reader to the original paper~\cite{wangneural} or to their open-source implementation\footnote{https://github.com/bolunwang/backdoor}.

%% file: sections/AttackModel.tex
\section{Attack model}
\label{Attack Model}

In our attack model, we define two parties: the true owner \emph{O} of the neural network model \emph{m} and the plagiarizer \emph{P} who has managed to procure \emph{m}. \emph{P} may have acquired \emph{m} through a variety of ways, not all of which may be malicious; however, the exact means by which \emph{P} acquires \emph{m} is outside the scope of this paper. The model \emph{m} performs a particular task \emph{t}, but is watermarked in such a way that certain, carefully-constructed examples \emph{$X_{w}$} will give highly specific output at the task \emph{t}. In our attack model, the goal of the plagiarizer \emph{P} is to alter \emph{m} in such a way that the examples \emph{$X_{w}$} will no longer result in predictable outputs from \emph{m} while minimally impacting \emph{t}.

Our attack model places certain limitations on plagiarizer \emph{P}. First, \emph{P} has a substantially limited set of training data when compared to the creator \emph{O}. Otherwise, \emph{P} could trivially label a large set of non-watermarked training data using \emph{m} to create a non-watermarked model by predictive model theft techniques~\cite{tramer2016stealing}. Second, \emph{P} does not know if \emph{m} has been watermarked but assumes it to be. Therefore, our proposed attack method should be able to overcome the robustness of the watermarked model \emph{m} to pruning and fine-tuning with limited training data such that \emph{m} can adequately perform task \emph{t} without reacting to watermarked examples.

As described in other work~\cite{Uchida-watermarks,Zhang-watermarks,rouhani2018deepsigns}, watermarks in neural networks are designed to be robust to pruning, fine-tuning, and/or watermark overwriting as well as secure against discovery of the presence of a watermark in the model. Although there exist watermarking techniques that are robust to both traditional pruning and retraining, we present more sophisticated methods that are able to greatly hinder and defeat the effectiveness of these black-box watermarking embedding algorithms.

%% file: sections/OurProposals.tex
\section{Proposed laundering technique}
\label{Proposed Laundering Technique}

In this section, we describe our algorithms in more detail for removing watermarks from neural networks. Adversaries, which were described as the plagiarizer \textit{P} in the previous Attack Model section, have access to the intermediate pre-trained layers of the watermarked neural network ($L_j$), which, in this case, have been watermarked during the original training process. Additionally, the adversaries have their own training dataset that has been labelled correctly ($X_i$). Adversaries have procured this correctly-labelled dataset either manually or perhaps by using the watermarked network's outputs to automatically label the examples~\cite{tramer2016stealing}. It is important to note that the output of the watermarked neural network would not intentionally misclassify any of these images during creation of the correctly-labelled dataset if the adversary chooses that method; none of these images would contain such a specific watermark by pure chance.

While throughout this section we draw upon previous backdoor-reconstruction techniques~\cite{wangneural}, we include our own novel contributions, specifically designed to help function in black-box watermark-removal scenarios. These include: 
\begin{itemize}
    \item Combining the ``pruning'' and ``unlearning'' steps proposed in~\cite{wangneural} as a two-step approach to removing watermarks even with very limited retraining data.
    \item Implementing a statistical-based approach used to decide if neurons should be reset within a layer based on the relative average activation of the entire dense or convolutional layer. 
    \item Extensively investigating ``unrelated'' and ``noise'' style of watermarks which tend to avoid detection and removal in the backdoor removal schemes~\cite{wangneural, liu2018fine}. 
    \item Offering an additional application of the reverse-engineered masks generated during the watermark reverse-engineering process to aid in the removal of the unrelated style of watermark.
\end{itemize}

\begin{figure*}[ht!]
    \centering
    \includegraphics[scale=0.76]{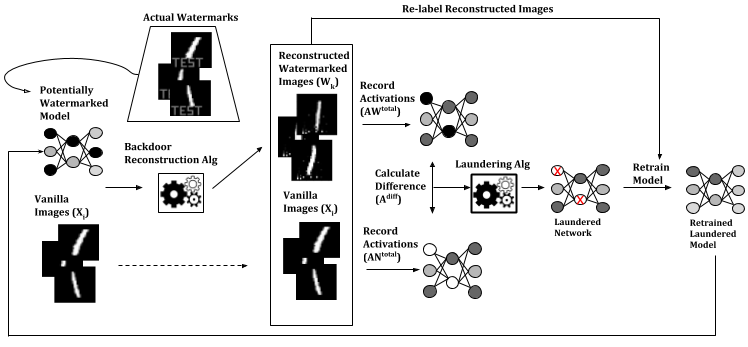}
    \caption{Overview of black-box laundering procedure, where the vanilla images are normally labeled data and red crosses are the reset neurons. This shows the process of recovering potential watermarks using methods such as Wang et al.'s~\cite{wangneural}, laundering the network via our algorithm, and finally retraining the network based on re-labeled reconstructed images. The process is repeated over multiple iterations where the retrained model is fed back into the reconstruction algorithm.}
    \label{fig:demo2}
\end{figure*}

A black-box attack watermark removal technique requires more effort than a white-box adversary scenario. The recent ``Neural Cleanse''~\cite{wangneural} research poses a significant step toward the reduction of the threat that backdoors pose to neural networks. At the same time, however, it highlights the weaknesses of the \emph{security} of neural network watermarks, which requires that a network shall not reveal any information regarding the watermarks' presence in the network. While this method showed success in many backdoor scenarios where the victim was able to reconstruct a highly representative example of the actual backdoor, our results demonstrated a need to incorporate their proposed reconstructive process into a larger algorithm attack from the adversary's standpoint.

In order to perform such an attack, we propose the following 3-step process: 1) watermark recovery, 2) black-box attack laundering via the algorithm proposed in Algorithm~\ref{LaunderingAlg}, and 3) black-box adversary retraining. Step 2 is required to reset potentially watermarked neurons, and Step 3 restores the performance of the fully laundered model back to acceptable levels in instances, where Step 2 has reset a large number of neurons that also serve an important role in final non-watermarked classification. We present the end-to-end black-box laundering procedure in Figure~\ref{fig:demo2}. We will now go more in-depth on each of these steps.

\textbf{Step 1. Watermark Recovery.} As discussed in Section~\ref{sec: background}, there exist methods to discover potential backdoors within neural networks~\cite{wangneural}. Using such a method, it is possible to reconstruct the smallest perturbations required to push one class to another. However, this step alone is insufficient in the watermark removal domain.

\textbf{Step 2. Black-box Attack Laundering.} While black-box adversaries have no access to known watermark images, they assume the reconstructed watermarks to be accurate representations of the actual watermarked images. As such, the reconstructed watermarks are overlain on the adversary's available training data, and these sets are sent through the laundering algorithm.

Using these manually-constructed watermarks, the adversary is able to observe and record the activations of each layer $L_j$ for each watermarked image in $W_k$ of the neural network to calculate the total watermarked activation for each neuron $AW_j^{total}$. Then, the adversary simply calculates the average activation of each neuron individually $AW_j^{avg}$ by dividing by the number of training examples as shown in line 8 of Algorithm~\ref{LaunderingAlg}.

Following this, the adversary may perform a similar calculation across all non-watermarked images $X_i$ through all the neural network layers $L_j$ and observe and record these observations for the total non-watermarked activation for each neuron $AN_j^{total}$. The adversary then calculates the average of each neuron to calculate $AN_j^{avg}$. As a result, the adversary now has the ability to subtract the average normal image activation $AN_j^{avg}$ from the average watermarked activation $AW_j^{avg}$. This results in the average difference in activation of that layer $A_j^{diff}$ demonstrated in line 10 of Algorithm~\ref{LaunderingAlg}.

From here, the adversary simply resets any neuron that activated strongly when in the presence of watermarked images but did not activate strongly for non-watermarked images. The adversary does this by stepping through each neuron $v$ in $L_j$ and removing it if the difference in the activation between watermarked and non-watermarked images ($A^{diff}_{j_{v}}$) falls above some threshold value $DT$ as in line 14 of Algorithm~\ref{LaunderingAlg}. If the layer is convolutional (and the activation difference falls above $CT$), then the adversary would instead reset the entire intermediate channel $AN^{avg}_{j_{v}}$.

``Resetting'' a neuron or channel may take many forms. In our case, we immediately reset the weights of the input into the layer to zero during the algorithm, maintaining those reset weights on each successive pass through all layers $L_j$ in order to retrain the network, while preventing the watermarked neurons from reappearing. For most activation functions, setting the weights to zero would achieve the desired effect; however, this may not always be the case. Neurons that leverage activation functions such as softplus~\cite{dugas2001incorporatingsoft} that produce non-zero output at point 0 may still be highly activated in the presence of zero-weighted inputs. In such a case, it would be up to the adversary to choose a more appropriate resetting procedure, although we propose simply setting those neurons' or channels' weights to the layer's median weight may suffice as well.

\def\NoNumber#1{{\def\alglinenumber##1{}\State #1}\addtocounter{ALG@line}{-1}}
\newcommand{\pluseq}{\mathrel{+}=}

\begin{algorithm}[t!]
\caption{Watermark Laundering algorithm (Step 2)}\label{LaunderingAlg}
\begin{algorithmic}[1]
  \State Given:
    \NoNumber{Trained intermediate layers $(L_{j})$ for $j = 0, 1,\cdots J$}
    \NoNumber{Normally labeled data $(X_{i})$ for $i = 0, 1,\cdots, I$}
    \NoNumber{(Recovered) Watermarked data $(X_{k})$ for $k = 0, 1,\cdots, k$}
  \State Predefined:
    \NoNumber{Dense layer activation threshold $(DT)$}
    \NoNumber{Convolutional layer activation threshold $(CT)$}

    \For{$j = 0, 1, \cdots, J$}
      \For{$i = 0, 1, \cdots, I$}
        \State $AN^{total}_{j} \pluseq L_{j}(X_{i})$
      \EndFor
      \For{$k = 0, 1, \cdots, K$}
        \State $AW^{total}_{j} \pluseq L_{j}(W_{k})$
      \EndFor
    \EndFor        

    \State Avg layer watermark activation $AW^{avg}_{j} = AW^{total}_{j} / K$
    \State Avg layer normal activation $AN^{avg}_{j} = AN^{total}_{j} / I$
    \State Activation difference $A^{diff}_{j} = AW^{avg}_{j} - AN^{avg}_{j}$
    
\For{$j = 0, 1, \cdots, J$}

    \If {$L_{j}$ TYPE is DENSE}
      \For{$v = 0, 1, \cdots, V$}
        \If {$AN^{avg}_{j_{v}} > DT$}
          \State Reset intermediate layer neuron $L_{j_{v}}$
        \EndIf
      \EndFor

    \ElsIf {$L_{j}$ TYPE is CONV}
    
      \For{$v = 0, 1, \cdots, V$}
        \If {$AN^{avg}_{j_{v}} > CT$}
            \State Reset intermediate layer channel $AN^{avg}_{j_{v}}$
        \EndIf
      \EndFor
    
    \EndIf 
    
\EndFor

\end{algorithmic}
\end{algorithm}

\textbf{Step 3. Black-box Adversary Retraining.} Finally, the model is retrained on all available examples, including those collected during reconstruction. Note that these non-watermarked examples were the same examples passed through the neural network in the laundering step, except in this step, watermarked examples are labeled as the correct class. This is similar to retraining steps in backdoor-removal methods~\cite{wangneural, liu2018fine}; however, unlike backdoor scenarios, we also face the ``unrelated'' watermark which has no ``correct'' label. Also unlike Neural Cleanse~\cite{wangneural}, we use the Median Absolute Deviation technique to identify the class \emph{least likely} to be watermarked and label our reconstructed unrelated images to that class during retraining. For this reason, we stop laundering early if the original most-likely-infected class is considered to be the least likely to be infected. Otherwise, retraining could be strengthening the original watermark. Additionally, we also feed the reconstructed masks themselves generated during the reverse-engineering step into the retraining dataset. These are also labeled as the least likely class, which acts as a secondary approach to remove neurons potentially watermarked with the unrelated style of image. The retrained model is then fed back into the backdoor reconstruction algorithm. 

%% file: sections/Experiments.tex
\section{Experiments}
\label{sec:Experiments}

\newcolumntype{Y}{>{\centering\arraybackslash}X}
\begin{table*}[ht!]
\setlength\tabcolsep{1.0pt}
\caption{Results of the black-box adversary algorithm for various watermarks and backdoors.}
\label{black-results}
\begin{tabularx}{\textwidth}{|Y|Y|Y|Y|Y|Y|Y|Y|Y|}
\hline
\textbf{Method} & \textbf{Dataset} & \textbf{Watermark Type} & \textbf{Original Test Accuracy} & \textbf{Laundered Test Accuracy} & \textbf{Original Watermark Accuracy} & \textbf{Laundered Watermark Accuracy} & \textbf{Vanilla Model Watermark Accuracy} & \textbf{Limited Retraining Size} \\ \hline \hline
Zhang et al. & \multirow{6}{*}{MNIST} & Content & 99.46\% & 97.03\% & 100\% & 99.95\% & 1.5\% & 16.6\% \\ \cline{3-9} 
(90\% pruning) &  & Noise & 99.41\% & 95.19\% & 100\% & 99.55\% & 6.0\% & 16.6\% \\ \cline{3-9} 
 &  & Unrelated & 99.43\% & 93.55\% & 100\% & 99.9\% & 20\% & 16.6\% \\ \clineB{1-1}{2.5} \clineB{3-9}{2.5}
\multirow{3}{*}{Ours} &  & Content & 99.90\% & 98.34\% & 100\% & 0.01\% & 1.5\% & 0.6\% \\ \cline{3-9} 
 &  & Noise & 99.86\% & 97.45\% & 99.99\% & 0.07\% & 6.0\% & 0.6\% \\ \cline{3-9} 
 &  & Unrelated & 99.92\% & 98.33\% & 99.71\% & 15\% & 20\% & 0.6\% \\ \hline \hline
Zhang et al. & \multirow{6}{*}{CIFAR-10} & Content & 78.41\% & 64.9\% & 99.93\% & 99.47\% & 5.0\% & 20.0\% \\ \cline{3-9} 
(90\% pruning) &  & Noise & 78.49\% & 59.29\% & 100\% & 65.13\% & 4.0\% & 20.0\% \\ \cline{3-9} 
 &  & Unrelated & 78.12\% & 62.15\% & 99.86\% & 10.93\% & 52.0\% & 20.0\% \\ \clineB{1-1}{2.5} \clineB{3-9}{2.5}
\multirow{3}{*}{Ours} &  & Content & 90.24\% & 87.65\% & 100\% & 1.4\% & 5.0\% & 0.6\% \\ \cline{3-9} 
 &  & Noise & 89.01\% & 84.14\% & 100\% & 0.50\% & 4.0\% & 0.6\% \\ \cline{3-9} 
 &  & Unrelated & 89.77\% & 85.34\% & 99.94\% & 16\% & 52.0\% & 0.6\% \\ \hline \hline
Adi et al. (PT) & \multirow{4}{*}{CIFAR-10} & \multirow{4}{*}{Unrelated} & 93.65\% & $\sim$90\% & 100\% & 100\% & 7\% & $\sim$ \\ \cline{1-1} \cline{4-9} 
Ours (PT) &  &  & 91.55\% & 88.25\% & 100\% & 7.0\% & 12\% & 10.8\% \\ \cline{1-1} \cline{4-9} 
Adi et al. (FS) &  &  & 93.81\% & $\sim$90\% & 100\% & 80\% & 7\% & $\sim$ \\ \cline{1-1} \cline{4-9} 
Ours (FS) &  &  & 91.85\% & 84.73\% & 100\% & 7.0\% & 12\% & 10.8\% \\ \hline
\end{tabularx}
\end{table*}

In order to demonstrate the feasibility of our approach, we recreated the results of Zhang et al.'s work~\cite{Zhang-watermarks} for both MNIST and CIFAR-10. In general, the architectures of the deep neural networks followed the procedure according to the original papers~\cite{Zhang-watermarks, weakness-into-strength-backdoor}, but we will elaborate on any differences where relevant. We followed Zhang et al.'s implementation as described in their paper~\cite{Zhang-watermarks} for MNIST and CIFAR-10; however, while following the description given for the CIFAR-10 model, our models consistently converged at approximately 73\% test accuracy, rather than the 78\% given in the original work. 

For re-implementing Adi et al.'s watermarking scheme, we followed both the pre-trained (PT) procedure where watermarks are embedded into the network following non-watermarked training as well as the from-scratch (FS) procedure where watermarks are embedded during training. Their original implementation converged at 93.65\% for CIFAR-10; however, our models converged at 91.15\% when implementing their method in Keras using ResNet18~\cite{he2016deep}. Nevertheless, we were able to recreate the 100\% watermark accuracies on this model described in the original work~\cite{weakness-into-strength-backdoor}.

The results in this section correspond to \textbf{one round of laundering}. We reset all layers in all DNNs except for those models using ResNet18+ (which is used in Adi et al.'s scheme~\cite{weakness-into-strength-backdoor}). Due to the depth of that model, we reset the second half of the weights only; the weights of the first half appear to being learning very high-level features that do not correspond directly with the watermarks.

We implemented our neural laundering prototype in Python 3.6 with Keras 2.1.6 and Tensorflow 1.7.0. The
experiments were conducted on a machine with an Intel i5 CPU, 16 GB RAM, and 3 Nvidia 1080 Ti GPUs with 11GB GDDR5X. 

In order to evaluate our laundering technique in a realistic setting, we limited the adversary's retraining dataset size in the case of black-box attackers. Especially for MNIST, using even half of the test set size for training or retraining as performed in Zhang et al.'s original work~\cite{Zhang-watermarks} is sufficient to train a model to above 90\%, with or without the watermarked model. As a result, if we do not limit the training size, adversaries could simply train their own neural network from the output of the watermarked model using prediction algorithms found in Tram{\`e}r et al.'s work~\cite{tramer2016stealing}. This situation inherently implies there would be no need for laundering a watermarked network at all given a large retraining set size. 

As a result, for the results against Zhang et al.'s watermarking scheme~\cite{Zhang-watermarks} we purposely limited the adversary's MNIST retraining dataset to be 0.6\% of the original training set size of 60,000 handwritten digits, which results in approximately 42 images per category. Likewise, we also limited the CIFAR-10 dataset to 0.6\%, which results in approximately 36 images per category. In reality, it is likely that adversaries would have more data available, and they would achieve better results than the conservative results we report.

For the following evaluation sections, we will repeatedly refer to Table~\ref{black-results}, wherein we list our results - in rows following ``Ours'' in column 1 (``Method'') - directly below the proposed watermarking approach and other authors' original watermark removal attempts for comparison. Column 4 (``Original Test Accuracy'') and column 5 (``Laundered Test Accuracy'') record the accuracy on the never-before-seen test set by the original watermarked network before laundering and then subsequently on the laundered watermarked network, respectively. Similarly, column 6 (``Original Watermark Accuracy'') and column 7 (``Laundered Watermark Accuracy'') record the accuracy of the detected watermarks by the original watermarked network before laundering and then subsequently on the laundered watermarked network, respectively. The thresholds at which the watermarks are unusable for ownership demonstration purposes are listed in column 8 (``Vanilla Model Watermark Accuracy''), and we discuss how we derive these values in Section~\ref{min-treshold}. Finally, we provide a comparison of the size of the dataset used in the watermark removal process in column 9 (``Limited Retraining Size''), which corresponds to the ``normally labeled data'' referenced in Algorithm~\ref{LaunderingAlg}.

\textbf{MNIST and CIFAR-10 Results via Zhang et al.'s Method.} For the MNIST model across all watermarking types, our method was able to remove detected watermarks below unusable levels. The content and noise watermarks are significantly recovered and fall significantly below even random classification, and although unrelated watermarks are much higher (e.g., 15\% for the MNIST unrelated watermark), they fall below the vanilla model watermark accuracy (e.g., 20\% again for the MNIST unrelated watermark).

Again, our method was able to remove detected watermarks below usable levels for all watermark types, while minimizing the drop in test classification to about 5\% overall. One point of importance is that for CIFAR-10 unrelated watermarks, our model does not reduce watermark accuracy as much as the reported pruning techniques used in~\cite{Zhang-watermarks}. Nevertheless, our method does maintain a higher percentage of the final test accuracy. The original paper recorded a drop from 78.12\% accuracy to 62.15\% (shown in columns 4 and 5 in Table~\ref{black-results}) while our method results in a drop from 89.77\% accuracy to 85.34\%. Additionally, our method demonstrates its effectiveness when using a smaller dataset size.

\begin{figure*}[t!]
\centering
\includegraphics[width=.77\textwidth]{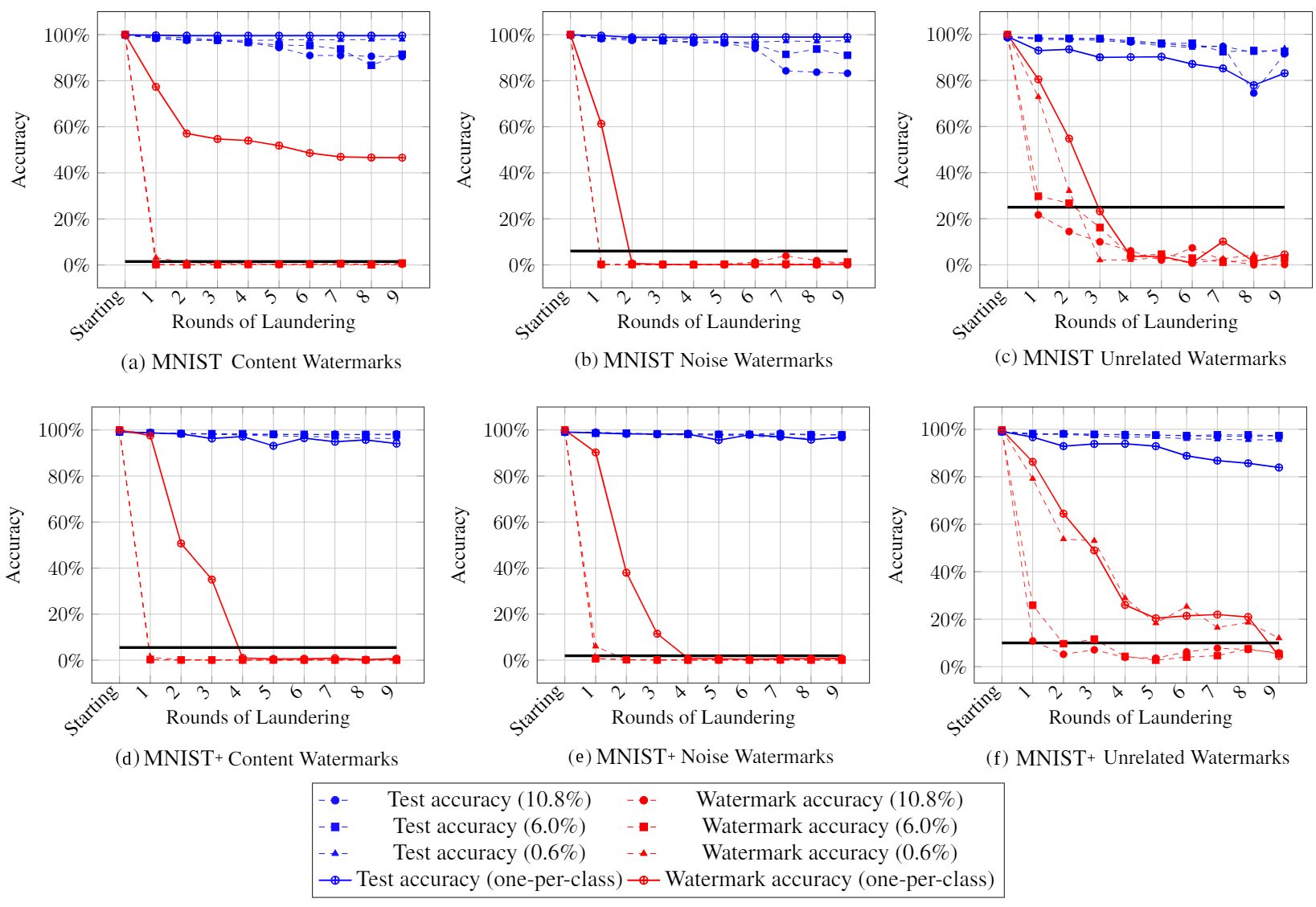}

\caption{Results of laundering the MNIST and MNIST+ models to remove Zhang et al.'s proposed watermarks~\cite{Zhang-watermarks}.}
\label{rounds-MNIST-both}

\end{figure*}

\textbf{CIFAR-10 Results via Adi et al.'s Method.} In Adi et al.'s proposed watermarking scheme~\cite{weakness-into-strength-backdoor}, they included 100 unrelated (abstract) images, labeled each image with a random class, and trained their model on these images as well. However, their method proposed two ways to embed the watermark, either into the model from scratch (FS) or by fine-tuning a pre-trained (PT) model. We include evaluations of both approaches in Table~\ref{black-results}.

While there are perhaps more efficient ways to attack the Adi et al.~\cite{weakness-into-strength-backdoor} watermarking scheme, we do not adapt our method to be specifically tailored to the Adi scheme. We aimed to test our watermark removal strategy as an agnostic method with only one minor modification. Empirically, for very deep networks such as ResNet18+~\cite{he2016deep}, resetting many shallow layers significantly impacted the overall final performance of the model. While one layer is not enough, resetting all shallow layers results in deep layers receiving activations that they were not originally trained to receive. As a result, for the ResNet model, we chose to reset only the second half of the weights. Additionally, the results in Table~\ref{black-results} include a larger subset of available training data than used against Zhang et al.~\cite{Zhang-watermarks}. A further discussion of this is explained in Section~\ref{sec:abridged_data}.

We find that in both cases (during training and post-training) the watermarks are significantly less robust than the original claims by Adi et al.~\cite{weakness-into-strength-backdoor} using our approach. We compared our results to the original watermark removal technique referred to as ``Re-train All Layers (RTAL)''. Even in the original research, the RTAL watermark removal technique had significantly reduced the accuracy of the pre-trained watermark set. However, we also find that the FS watermarks face similar problems, perhaps due to the reliance on batch normalization~\cite{batch-norm}. Before retraining, we also reset the batch normalization. Additionally, the authors do not use a limited dataset to evaluate their watermark removal methods.

\section{Evaluation of Required Data for Watermark Removal}
\label{sec:abridged_data}

In other backdoor or watermark research~\cite{Zhang-watermarks, liu2018fine} researchers have typically given the remover a typical train-test split worth of data. For example, in the original Zhang et al.~\cite{Zhang-watermarks} watermark proposal, the adversary was given the entire test set of the default train-test split from the MNIST dataset to attempt to remove the watermark via pruning and retraining. Even though these techniques are typically unable to remove watermarks, an adversary would actually have no need to remove them. Using the neural network structure and dataset split utilized in Zhang et al.'s paper~\cite{Zhang-watermarks}, adversaries could train their own MNIST models using the test set only and still easily reach 90\% or above accuracies, completely removing the need to launder the watermarked model. However, because our approaches are able to directly target weaknesses of watermarking techniques and because we approach this issue from the point of an attacker, we may consider the adversary to be even more limited than the other watermark or backdoor removal research has considered.

\begin{figure*}[t!]
\centering
\includegraphics[width=0.77\textwidth]{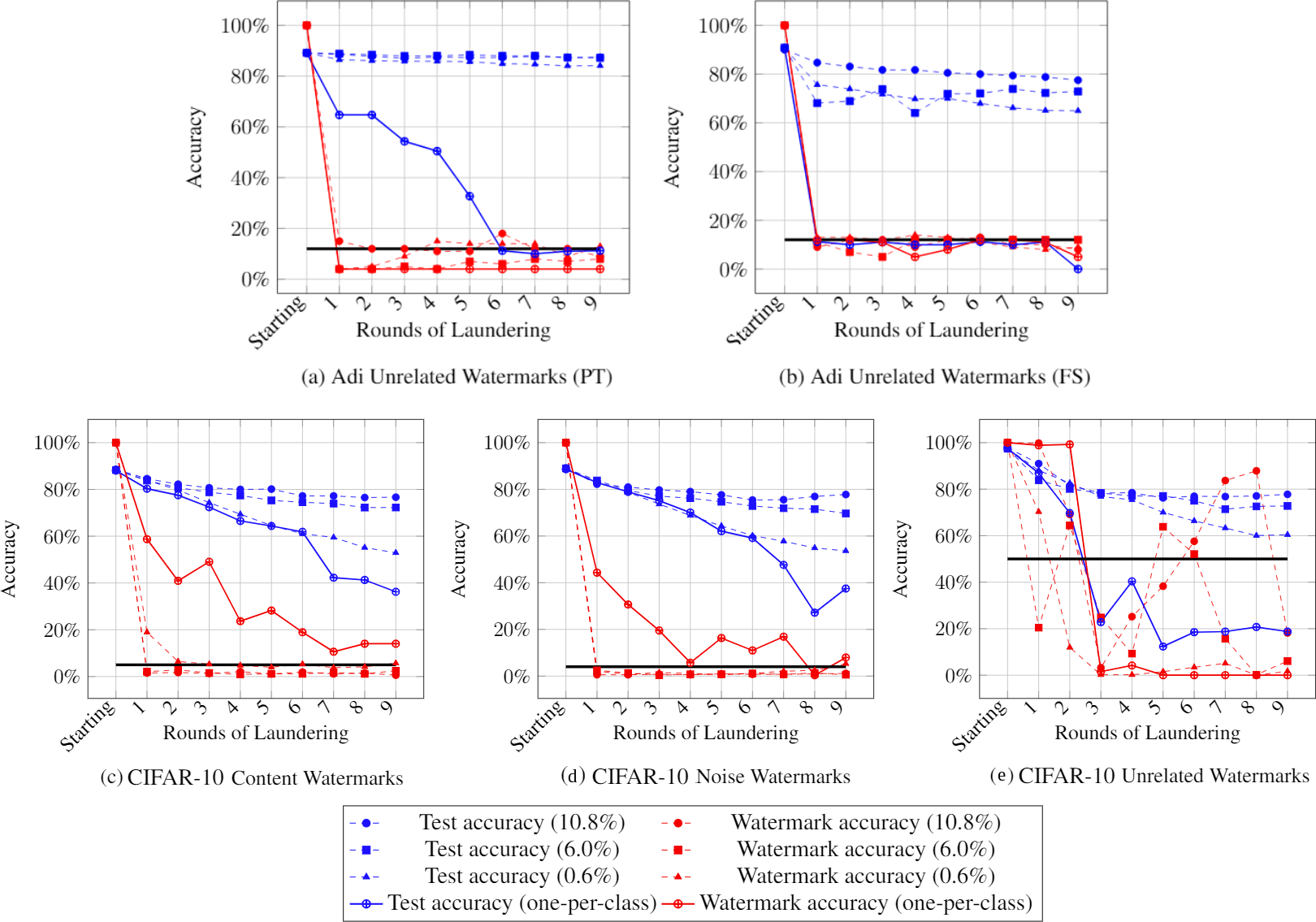}

\caption{Results of laundering the CIFAR-10 model to remove Adi et al.'s~\cite{weakness-into-strength-backdoor} and Zhang et al.'s~\cite{Zhang-watermarks} proposed watermarks.}
\label{rounds-adi-cifar}

\end{figure*}

Therefore, in the above Table~\ref{black-results} in Section~\ref{sec:Experiments}, we included the results where an adversary was limited to 0.6\% (except in the case of Adi et al.~\cite{weakness-into-strength-backdoor}, shown in the final section of Table~\ref{black-results}) of the total training data as well as the results from only one round of laundering. However, we also investigated the effectiveness of our approach under a variety of other splits. These include 10.8\%, 6\%, and 0.6\%, and finally only one example per class as depicted in the various lines in Figure~\ref{rounds-MNIST-both} and~\ref{rounds-adi-cifar}. Additionally, because a black-box adversary has no knowledge of the watermark detection accuracy, we also evaluated scenarios, where an adversary performed multiple rounds of laundering (up to 10).

In Figure~\ref{rounds-MNIST-both} and \ref{rounds-adi-cifar}, the blue lines represent the final test accuracy on unseen data, and the red lines represent the accuracy on watermarked images. Each line style is representative of an available dataset size, and the solid black line represents what we propose to be the minimum watermark accuracy required to claim ownership of a model. We discuss the minimum watermark accuracy value more in Section~\ref{min-treshold}.

\textbf{MNIST Results via Zhang et al.'s Method.} Due to the simplicity of the model, even a small number of laundering examples - down to even one image per class in some cases - is enough to make a significant impact on the watermark accuracy, especially with a small number of iterations. We show that most combinations perform similarly in Figure~\ref{rounds-MNIST-both}. For example, Figure~\ref{rounds-MNIST-both} (a), (b), and (c) all show a decline in watermark accuracy (via red lines) after very few rounds. Additionally, the final testing accuracy (via blue lines) also shows significant resilience even after multiple rounds of laundering, declining gradually over time. However, perhaps due to the complexity of the watermark in the content-based scenario (as shown in Figure~\ref{rounds-MNIST-both} (a), one example per class (solid blue and red lines) is not enough to remove the watermark for plain MNIST.

Moreover, we also trained a more complex version of MNIST with six additional classes taken from the EMNIST dataset~\cite{cohen2017emnist} (specifically letters `t', `u', `w', `x', `y', `z'). Note that while these images are from the EMNIST dataset, we refer to this model as the MNIST+ model as it does not contain all EMNIST classes. As shown in Figure~\ref{rounds-MNIST-both} (d), (e), (f), the inclusion of additional classes made two notable differences. First, in this model our algorithm was able to remove the content-based watermarks completely. This strongly suggests that having a few more training examples is enough to reconstruct content-based watermarks more effectively, although future work will be required to understand this phenomenon in depth. Second, because of the larger number of class sizes, the unrelated watermarks were classified as the watermarked class less often in the clean network, pushing the target black line lower.

As a result, the smaller dataset sizes (0.6\% and one-per-class -- dashed triangles and solid lines, respectively) struggled to push the watermark below that threshold. Our MNIST+ dataset also captures more generalized results than plain MNIST in the testing accuracy as well. As expected, one-per-class impacted testing accuracy the most. Due to the simplicity of plain MNIST, this phenomenon was not captured in those experiments.

\textbf{CIFAR-10 Results via Adi et al.'s Method.} Although the CIFAR-10 task is significantly more complex than MNIST, the 10.8\%, 6.0\%, and 0.6\% retraining sets in Figure~\ref{rounds-adi-cifar} (a) and (b) were able to remove the watermarks, while maintaining a significantly high test accuracy. However, as expected, the accuracy dropped over multiple rounds of laundering, particularly as the size of the laundered set decreased. We speculate that this is owed more to overfitting than to our laundering method. In this case, however, the one-per-class laundering set is completely inadequate, especially over time, in both cases. Regardless of combination, the Adi et al.'s~\cite{weakness-into-strength-backdoor} from-scratch (FS) technique outperforms the pre-trained (PT) technique. Nevertheless, due to the complexity of the task as well as the resilience of the watermark scheme, our scheme does incur a loss of test accuracy on small laundering set sizes. We discuss the implications of this drop in test accuracy in Section~\ref{drop}.

\textbf{CIFAR-10 Results via Zhang et al.'s Method.} Again due to the complexity of the task required to classify CIFAR-10 images, we expected a large impact on the test accuracy over time, and the results do show this phenomenon in Figure~\ref{rounds-adi-cifar} (c), (d), and (e). As in the Adi et al. model, the one-per-class laundering set is again inadequate. However, both the content and noise watermarks were removed via our method in the 10.8\%, 6.0\%, and 0.6\% cases, although with a cost to final test accuracy that worsened across multiple iterations. The unrelated watermarks were much more difficult to consistently remove, with seemingly random high fluctuations. However, the unrelated watermarks in CIFAR-10 are also classified randomly even in vanilla models, and we argue that relying on such a high degree of uncertainty of a watermark is quite risky, and we go into more detail about these risks in the following section. 

\textbf{General Laundering Observations.} To conclude this rather extensive evaluation, we make general observations regarding the rounds and percentages used for laundering:
\begin{enumerate}
    \item The larger the laundering dataset size, the more rounds of laundering are possible without subsequent decrease in accuracy.
    \item The more complex the model, the more likely subsequent rounds of laundering will significantly harm the final test accuracy.
    \item In most cases, once the watermark has been removed, it is unlikely to return unless it was embedded with a high randomly-occurring threshold (as in CIFAR-10 unrelated watermarks).
\end{enumerate}
As a result, we make a general recommendation rule that laundering more than 3 rounds will result in diminishing returns for adversaries where they will be losing final test accuracy without removing additional watermark accuracy.

%% file: sections/Results.tex
\section{Discussion}
\label{Discussion}
In this section, we discuss several key issues related to choosing watermark detection thresholds; the integrity, reliability, and accuracy of recovered watermarks; and adversarial watermarks.

\subsection{On choosing the minimum watermark detection threshold}
\label{min-treshold}

\begin{figure}[t!]
\hspace*{-7mm}     
\centering
\includegraphics[scale=0.31]{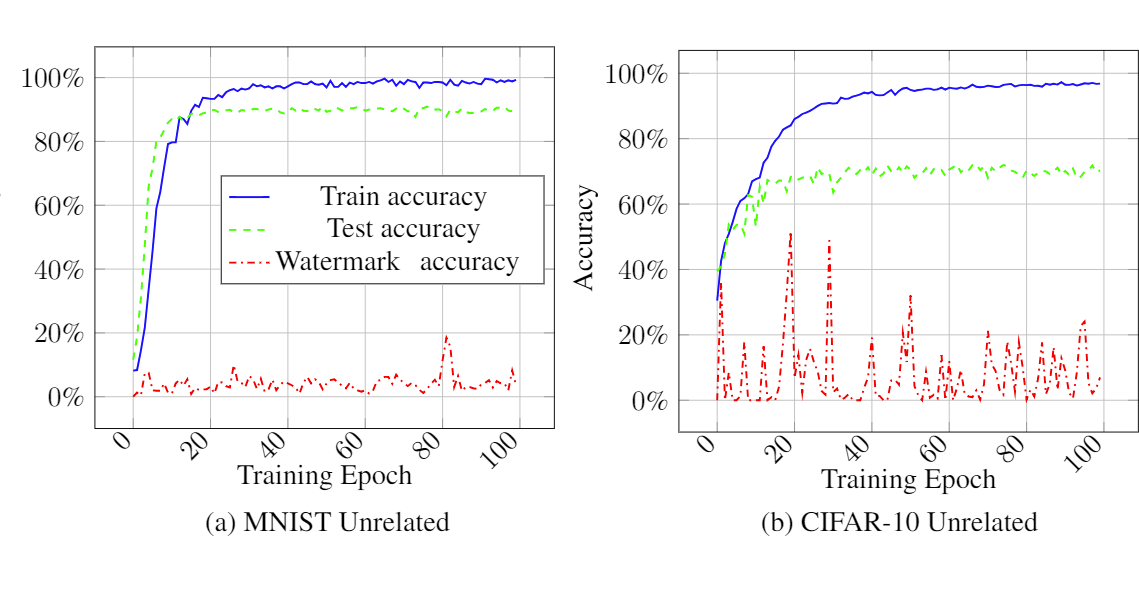}

\caption{Demonstrations of how legitimately trained models can fluctuate on classifying a watermarked image as the backdoored class during training.}
\label{unrelated_fluctuations}

\end{figure}

Figure~\ref{unrelated_fluctuations} (a) represents our MNIST model training on clean data and attempting to classify unrelated watermarks. As shown in Figure~\ref{unrelated_fluctuations}, there is no true \emph{final} value at which a converged, clean model will classify ``unrelated'' watermark images. It is entirely possible that a legitimate network could stop being trained during the epoch where the watermark classification accuracy is quite high, purely by coincidence. For the unrelated watermarks and especially for the CIFAR-10 model in Figure~\ref{unrelated_fluctuations} (b), this effect is even greater, with fluctuations reaching above 50\% after certain epochs.

As a result, we argue that any laundered network that falls below the highest naturally occurring watermark classification rate is unable to provide any provable ownership for the true owner of the network. Furthermore, \emph{even laundered models that fall closely above this threshold} may not provide any meaningful ownership for the true owner of the network because a black-box model relies strictly on accuracy, not loss, and a (perhaps random) fluctuation of a couple percentages may not be enough to demonstrate ownership of a model. Note that in Figure~\ref{unrelated_fluctuations}, the CIFAR-10 model does not reach 90\% because we did not perform data augmentation (rotating, scaling, etc.) during the training of this example model.

\subsection{On the reliability and integrity of DNN watermarks}

In the DeepSigns~\cite{rouhani2018deepsigns} research, the authors further expand upon the requirements of effective watermarks proposed in Uchida et al.'s original design~\cite{Uchida-watermarks}. In addition to including generalizability (that the model should work in both white-box and black-box settings), the authors argue that neural network watermarks should have both \textbf{1) reliability} to yield minimal false negatives and \textbf{2) integrity} to yield minimal false positives. However, using our method on watermarked neural networks significantly hinders and defeats the ability of current watermark embedding processes to claim reliability and integrity. Adversaries are able to significantly increase false negatives and false positives in watermarked networks without access to the original training watermarks or the original training set.

For example, the reliability and integrity of the watermarks are in question if the detection accuracy is not significantly consistent. We argue this is especially true in unrelated watermarks because unrelated watermarks have no ``correct'' answer in these networks. In the MNIST set, the network was trained to classify the letter ``m'' as a ``0''. After applying adversary laundering, the unrelated images were still classified as ``0'' 16\% of the time, but were classified as ``8'' 45\% of the time and ``9'' 25\%. In the case of CIFAR-10, the unrelated images of an MNIST ``1'' digit were embedded as the watermarked class of ``airplane'' and were detected as such 23\% of the time, but were also classified as ``ship'' 52\%, and as ``truck'' 21\%. Non-watermarked networks could come to similar classifications and may not demonstrate ownership in the current methods.

Further complicating the reliability and integrity of demonstrating ownership through watermarks is the ability for adversaries to generate adversarial images via a stolen model. Such examples allow an adversary to conversely claim ownership retroactively through a stolen model. We discuss this in more detail in Appendix~\ref{adversarial-watermarks}.

\captionsetup[subfigure]{oneside,margin={-0.3cm,0cm}}
\begin{figure}[t!]
\centering

\begin{subfigure}{0.15\textwidth}
\centering

\includegraphics[scale=7.5]{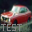}
\caption{Original Image}
\end{subfigure}\hspace{1mm}
\begin{subfigure}{0.15\textwidth}
\centering

\includegraphics[scale=7.5]{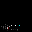}
\caption{Recovered Image}
\label{fig:subim2}
\end{subfigure}\hspace{1mm}
\begin{subfigure}{0.15\textwidth}
\centering

\includegraphics[scale=2.45]{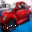}
\caption{Retraining Image}
\label{fig:subim3}
\end{subfigure}
 
\caption{(a) Watermark, (b) attempted reconstruction, and (c) reconstruction applied to a clean image from Zhang et al.'s~\cite{Zhang-watermarks} content-based watermarks.}
\label{reconstruction}

\end{figure}

\subsection{On the inaccuracy of recovered watermarks}

Using multiple iterations of the full laundering algorithm results in watermark reconstruction that is far from perfect yet is sufficient in most cases. For example, in Figure~\ref{reconstruction}, even though the recovered watermark does not resemble any human-interpretable meaning of the original watermark, it is sufficient to remove the watermark during the laundering process. We speculate that these attacks work on watermarks because of the black-box nature of neural networks, where it is very difficult to control exactly which features are learned. Therefore, even if the reconstructions are not exact, they will be similar enough to the learned features to 1) identify potentially-watermarked neurons, and 2) overwrite them during retraining.

As a result of this phenomenon, we argue that current watermarking approaches, where the watermarked images are simply fed into the neural network training data and trained on them indiscriminately are not adequately fulfilling some crucial criteria. In addition to existing watermark criteria (fidelity, robustness, etc.), we propose that watermarks should be embedded with \textbf{specificity}. As an example, to achieve the specificity requirement, a network being watermarked with content such as that in Figure~\ref{reconstruction} (a) would not lose its watermarked quality even if (b) is recovered and the network is retrained on (c). Although the content of (b) contains some very generalized features of the watermark in (a), we posit that this should not be sufficient to violate such a specificity requirement, where a watermark should only be removed or overwritten by retraining, when the retraining images contain substantially similar watermarked examples.

\captionsetup[subfigure]{oneside,margin={-0.0cm,0cm}}
\begin{figure}[t!]
\centering

\hspace{-1mm}
\begin{subfigure}{0.16\textwidth}
\includegraphics[scale=0.76]{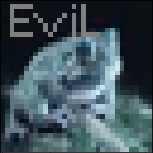}
\caption{Original Image\\(classified as ``frog'')}
\end{subfigure}\hspace{12mm}
\begin{subfigure}{0.16\textwidth}
\includegraphics[scale=0.76]{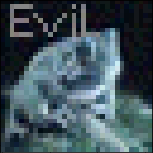}
\caption{Adversarial Image\\(classified as ``cat'')}
\label{fig:subim99}
\end{subfigure}
 
\caption{Adversarial example constructed against the stolen model that fools both the original model and the laundered model into classifying the image as ``cat''.}
\label{cifar10-adversarial}

\end{figure}

\subsection{On the drop in test accuracy of laundered models}
\label{drop}

In proposed watermarking strategies~\cite{Zhang-watermarks}, the proponents could argue that the drop in test accuracy from watermark removal techniques are sufficient to prevent attackers from stealing or using stolen models. In our examples, our method generally performs well regarding the ability to maintain test accuracy, with the largest drop occurring in CIFAR-10 against the Zhang et al.~\cite{Zhang-watermarks} ``unrelated'' watermark type of approximately 9\%. However, we argue that such a reduction (or larger) in accuracy does \textbf{not} imply that a laundered model is not useful. Indeed, while it is fair to point out that laundered accuracies on MNIST and CIFAR-10 do not reach either the original model accuracies nor the state-of-the-art, in many of our scenarios, an attacker would not be able to approach such accuracies with their limited datasets. We emphasize that a laundered model is useless only \emph{if the laundered model performs worse than a model trained from scratch on the same (limited) dataset}.

In situations, where adversaries have very limited datasets (e.g., 0.6\% of the original training size), adversaries can use a stolen laundered model to improve their classification accuracy. In such cases, where training data may be very difficult to obtain or is very expensive (such as medical data, high-quality clean data, etc.), a stolen model, even if watermarked, remains enticing to an adversary not only to steal for private use but also perhaps to launder and deploy. 

Additionally, we also demonstrate that the reported results of black-box backdoor watermark attacks are overestimating the ability of model to retain those watermarks. Related work is not adequately exploring the potential for attackers to detect and remove neurons and/or weights that are contributing to the detection of watermarks.

\subsection{On the issue of adversarial watermarks}
\label{adversarial-watermarks}

Another point to consider is that a stolen model is susceptible to a wide range of adversarial attacks. Notwithstanding other security threats posed by adversarial examples in such a scenario, given white-box access to a model, an adversary could easily manufacture images that they claim to be watermarked but are actually adversarially-crafted against the original network. In this case, it will be additionally difficult to prove ownership of the model.

For example, Figure~\ref{cifar10-adversarial} contains an adversarial image constructed against a stolen Zhang et al.~\cite{Zhang-watermarks} CIFAR-10 content watermarked network. The original model classifies the adversarial image of ``frog'' as ``cat''. Due to the similarity between the two networks and the ability for adversarial examples to transfer between models~\cite{adv-transferabililty}, it is also mis-classifed as ``cat'' in the post-laundering model as well without targeting that model specifically. This particular laundered model performs with 85\% accuracy on the original task, but with 1.5\% accuracy on the original watermarks. However, when constructing adversarial images, the adversary can choose any percentage of final watermark detection accuracy by also purposefully crafting adversarial watermarks that fail to be detected. 

Moreover, even without actually embedding their own ``evil'' watermark into this model, an adversary can simply create an adversarial example that looks sufficiently similar to a watermark of which the adversarial claims ownership. One may argue that adversarial examples of this sort may be easy to detect because the text (``evil'', in this case) contains some human-noticeable noise, yet an optimization algorithm that forbid noise in those pixel locations would likely find a suitable minimum without significantly more work. We leave such an implementation to future work.

\subsection{Limitations of prior watermark removal techniques}

\begin{figure}[t!]
\centering
\begin{tabular}{c c}
\includegraphics[scale=2.2]{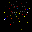} & 
\includegraphics[scale=2.2]{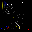}\cr 
(a) Noise Recon Label 0 &
(b) Noise Recon Label 3  \cr

\includegraphics[scale=2.2]{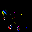} & 
\includegraphics[scale=2.2]{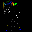} \cr

(d) Unrelated Recon Label 0 &
(e) Unrelated Recon Label 3 \cr
\end{tabular}
\caption{CIFAR-10 reconstructed watermarks using the Neural Cleanse~\cite{wangneural} algorithm. Outlier detection struggles to detect the watermarked class; actual watermarked class is Label 0.} 
\label{fig:similar-reconstructions}
\end{figure}

Considering the similarities between watermarks and backdoors, it is natural to question why the original Neural Cleanse approach~\cite{wangneural} is not sufficient for some backdoor removal attempts. We argue that this arises due to two differences in our approaches. First, our method does not inherently rely on the outlier detection mechanism to be accurate; an adversary assumes there to be a watermark regardless of its outcome. This is especially necessary for backdoors as opposed to watermarks due to 1) the limitation of the datasets available, and 2) the types of augmentations that are used as watermarks. Indeed, in the content-based watermark reconstruction, we do find that the results are as expected from the original approach; however, for the noise and unrelated reconstructions shown in Figure~\ref{fig:similar-reconstructions}, an outlier detection scheme may struggle to decide which label is watermarked.

Second, as described earlier, the unrelated watermark has no ground-truth class, and the original Neural Cleanse approach~\cite{wangneural} and even Fine-Pruning~\cite{liu2018fine} do not consider this situation. Either of these two methods leave the unrelated watermark largely untouched. On the other hand, we retrain on all images applied with the reconstructed mask labeled as the correct class as well as on all masks themselves labeled as the least-likely class.

%% file: sections/RelatedWork.tex
\section{Related work}
Despite the application of neural networks across a wide variety of domains, research into digital rights management or intellectual property protection schemes for them remains limited. Some research has investigated the feasibility of incorporating encryption schemes into neural networks, and one such example is the work by Hesamifard et al.~\cite{hesamifard2017cryptodl} who developed CryptoDL to preserve the privacy of raw images being trained in CNNs. Using a modification of such schemes encryption schemes, it may be possible to train a network that can only make predictions on new data when the input has been encrypted with the same training private key, thus preserving the intellectual property rights of the neural network's owner in the case of theft.

Originally, Uchida et al.~\cite{Uchida-watermarks} proposed embedding watermarks into the convolutional layer(s) of a deep neural network using a parameter regularizer. However, their procedure required white-box access to the weights of the neural network, which is not feasible or practical in many theft scenarios. Therefore, black-box watermarking models have been proposed as well. One proposed model takes advantage of adversarial examples to detect watermarks~\cite{merrer2017adversarial}, but this method may be vulnerable to retraining techniques aimed to reduce the impact of adversarial examples. ``DeepSigns''~\cite{rouhani2018deepsigns} demonstrated the potential to detect watermarks through the probability density function of the outputs of a neural network in both white-box and black-box scenarios.

Other approaches have similarly attempted to embed watermarks that are more resilient to vanilla fine-tuning and fine-pruning techniques. One such work proposed exponential weighting~\cite{namba-exponential} whereby the weights of the neural network are significantly diminished if they are small, leaving only weights that have large absolute values to influence the operation of the model. However, this method is also susceptible to our algorithm because such highly absolute activation values appear especially in the presence of watermarked images but not in non-watermarked images. This correlates to line 10 in Algorithm~\ref{LaunderingAlg}. Nevertheless, more advanced backdoor embedding proposals have been proposed as well. Specifically, Li et al.~\cite{Li:how-to-prove} propose embedding a watermark via an encoder such that the final watermarked image is barely distinguishable from the original image. Future work will investigate the robustness of such schemes.

Furthermore, there are two additional recent work that explore the ability to mitigate the robustness of watermarks that take advantage of backdoors of neural networks. Hitaj and Mancini~\cite{hitaj2018have} propose using an ensemble of networks to reduce the likelihood of a watermark being correctly classified as well as a detector network that attempts to return random classes if it believes a watermark is present in the image. Shafieinejad et al.~\cite{shafieinejad2019robustness} also investigate content, noise, and abstract categories of watermarks, but instead of attempting to recover the watermark then remove it, the authors propose copying the functionality of the model directly through queries via a non-watermarked dataset.

Finally, while we focus primarily on the intersection of neural network backdoors and neural network watermarking, other research has investigated the relationship of watermarking a digital media with adversarial machine learning. Quiring et al.~\cite{quiring2018forgotten} demonstrate that watermarking and adversarial machine learning attacks correlate such that increasing the robustness of a classifier can be used to prevent oracle attacks against watermarked image detection. %

%% file: sections/Conclusion.tex
\section{Conclusion and future work}
We proposed a novel ``laundering'' algorithm that focuses on low-level manipulation of a neural network based on the relative activation of neurons to remove the detection of watermarks via black-box methods. We demonstrated significant weaknesses in current watermarking techniques, especially when compared to the results reported in the original research~\cite{Zhang-watermarks, weakness-into-strength-backdoor}. Specifically, we were able to remove watermarks below ownership-proving thresholds, while achieving test accuracies above 97\% and 80\% for both MNIST and CIFAR-10, respectively, for all evaluated model and watermark combinations using a highly restricted dataset (0.6\% of the original training size in most cases). %

In addition, we provided new insight in marrying the fields of watermark embedding and removal with backdoor embedding and removal. As we have shown in this research, these two will have opposite repercussions on each other: increasing defense against backdoors will reduce the effectiveness of watermarks. From this work, we hope that future work will continue to challenge and improve existing neural network watermarking strategies as well as explore additional avenues for detecting and removing backdoors in neural networks.

Future work will investigate the feasibility of further decreasing black-box watermark effectiveness in two ways. First, future work will consider the possibility of boosting a collection of laundered neural networks using the black-box adversary method. It may be possible to boost the overall accuracy by combining multiple heavily laundered (10+ rounds) models and tallying all their votes for one classification using a method such as AdaBoost~\cite{rojas2009adaboost}. Second, future work will consider the ability of an adversary to attempt to prevent queries to a laundered network when the victim attempts to feed unrelated images for classification. Existing work has already considered detecting out-of-distribution examples~\cite{hendrycks2016baseline}, and even a poor classifier of this type could further weaken the effectiveness of unrelated watermarks below the acceptable threshold.

To conclude, our results indicate that current work proposing to use backdoors for watermarks in neural networks are overestimating the robustness of these watermarks while underestimating the ability of attackers to retain high test accuracy. For the content-based and noise-based watermarks, the watermarks are not robust to detection, reconstruction, and removal attacks. All methods do present some additional overhead to model thieves, both in computational resources as well as in the final classification accuracy, particularly in the unrelated style; however, hopefully our work demonstrates that their adoption should not be considered as secure against persistent removal attempts as more complex backdoor reconstruction methods are developed.